\documentclass[]{mn2e}
\usepackage{psfig, epsf, epsfig}
\title[Primordial pollution of GCs]{Primordial pollution of globular clusters
within their host dwarfs embedded in dark matter
halos at high redshifts}
\author[Kenji Bekki]
       {Kenji Bekki${}^1$\thanks{E-mail: bekki@bat.phys.unsw.edu.au} \\
        ${}^1$School of Physics, University of New South Wales,
              Sydney 2052, NSW, Australia}
  
\begin{document}

\date{Accepted, Received 2005 May 13; in original form }

\pagerange{\pageref{firstpage}--\pageref{lastpage}} \pubyear{2005}

\maketitle

\label{firstpage}

\begin{abstract}

Recent observational studies have revealed
star-to-star  abundance inhomogeneity among light elements
(e.g., C, N, O, Na, and Al) of stars
on the main sequence in the Galactic globular clusters (GCs). 
One of promising interpretations for this result
is that the observed abundance inhomogeneity is due to 
the second generation of
stars formed from ejecta of the first generation  of
evolved  stars (e.g., AGB stars) within GCs.
However it remains unclear whether and how this primordial pollution can 
occur within GCs. 
We here propose a new scenario in which primordial pollution of GCs is highly
likely to occur if GCs are located
in the central regions of high redshift dark matter
subhalos that can host low-mass dwarf galaxies.
In this scenario,  gas ejected from 
the first generation of stars of GCs 
can be effectively trapped in the deep gravitational
potential of their host halos
and consequently can be consumed for the formation of 
the second generation of stars without losing a significant
amount of gas by ram pressure stripping of interstellar 
and intergalactic medium. 
During merging of these halos with the proto-Galaxy,
the halos are completely destroyed owing to the strong
tidal field of the Galaxy. The self-polluted GCs located
initially in the central regions of the halos can survive
from tidal destruction owing to their compactness
and finally become the Galactic
halo GCs.
In this scenario, ejecta of field stars surrounding the central
GCs can be also converted into stars within their host dwarfs
and finally become the
second generation of stars of GCs. 
We also discuss the origin of the difference in the degree
of abundance inhomogeneity between different GCs, such
as $\omega$ Centauri and NGC 6752, in terms of
the difference in physical properties between host
halos from which GC originate.

\end{abstract}


\section{Introduction}

The origin of star-to-star abundance inhomogeneity observed in
the Galactic globular clusters (GCs) has long been
discussed based mainly on the following two working hypotheses:
The primordial hypothesis and the mixing one
(e.g., Cottrell \& Da Costa 1981; Freeman \& Norris 1981;
Smith 1987; Suntzeff 1993; Kraft 1994; Gratton et al. 2004).
The first hypothesis is that the observed inhomogeneity 
is due to the second generation of stars that 
were formed from  gas ejected from
the first generation of evolved stars (e.g., AGB stars) of GCs
(``primordial pollution''  
scenario, e.g.,  Cottrell \& Da Costa 1981). 
The second  is that the observed chemical inhomogeneity
of GCs can result from {\it internal processes} of stars,
such as dredge-up of CN-processed material from 
inner hydrogen-burning regions (e.g., Smith 1987; Kraft 1994).
Recent observational studies of stellar abundance
of some Galactic GCs have revealed
star-to-star  abundance inhomogeneity among less evolved stars
on the main sequence,
where deep mixing of chemical components are highly unlikely
(e.g., Cannon et al. 1998 for 47 Tuc).
These studies accordingly suggested
that the  primordial pollution scenario is more promising
than the mixing (or evolutionary) scenario
(Da Costa et al. 2004; Gratton 2004 for a recent review).

One of key questions related to the
primordial pollution scenario is whether and how ejecta mainly from
AGB stars of the first generation of stars
can be effectively trapped  within GCs and consequently
converted into the second  generation of stars. 
Frank \& Gisler (1976) showed that gas of GCs can 
be efficiently stripped by ram pressure of the Galactic halo
gas for most GCs.
Smith (1996) suggested that the stellar ejecta from 
the first generation of stars 
are likely to be lost entirely from GCs with small binding 
energies through energetic outflow of intracluster wind.
Gnedin et al. (2002) demonstrated that
$\omega$ Cen  could  not enrich itself with heavy elements
of AGB stars owing to efficient ram pressure stripping
of the Galactic interstellar medium (ISM), if it formed and 
evolved in isolation.
These previous studies thus appear to suggest
that primordial pollution is not likely 
within  GCs evolving in isolation,
though  numerical attempts 
have not yet been made to investigate 
the details of primordial pollution processes
within GCs.

Recent numerical simulations have suggested that
GCs can be formed in the central regions of dwarf galaxies
embedded by low-mass dark matter halos 
(e.g., Bromm \& Clarke 2002; Mashchenko \& Sills 2005).
Both observational and theoretical studies of GCs 
suggested that 
massive globular clusters such as $\omega$ Cen and G1
were previously 
stellar nuclei (or nuclear star clusters)
of nucleated dwarf galaxies 
(e.g., Freeman 1993; Bekki \& Freeman 2003; Bekki \& Chiba 2004).
Owing to the deeper gravitational potential in the central
regions of dark matter halos,
primordial pollution processes can be quite different
between GCs evolving in isolation and those in the nuclear regions 
of their host halos.
Thus it is quite important and timely to discuss
whether and how primordial pollution of GCs can proceed 
{\it if GCs are within the central regions of low-mass dark matter halos
that can host dwarf galaxy populations.}

The purpose of this Letter is to propose
a new scenario in which primordial pollution can proceed
very efficiently in GCs that are located in nuclear regions
of their host halos at high redshifts.
In this scenario, primordial pollution can proceed more efficiently
in the nuclear GCs
than in those evolving in isolation,
because the ejecta of AGB stars are more effectively
trapped in GCs (without being significantly lost through
ram pressure stripping and energetic outflow of evolved stars) 
owing to the deeper gravitational potential of
their host halos.
The nuclear, self-polluted GCs can appear as 
the Galactic halo GCs when the host halos merge with the proto-Galaxy
and are subsequently destroyed by the strong tidal field of the Galaxy
(i.e., when field stars and dark matter halos surrounding the GCs
are all removed by the tidal stripping).

We firstly show the three advantages of this scenario in
explaining why ejecta of the first generation of stars can be
effectively trapped and converted into stars.
Then we discuss the origin of the difference in the degree of
abundance inhomogeneity between GCs in the Galaxy
in the context of the proposed chemical pollution scenario.
Since many authors have already discussed advantages and disadvantages
of the primordial pollution (or enrichment) processes {\it within 
isolated proto-GC clouds and  GCs}
(e.g., Cayrel 1986; 
Parmentier \& Gilmore 2001;
Thoul et al. 2002;
Recchi \& Danziger 2005), 
we here do not intend to discuss these points.

\begin{figure}
\psfig{file=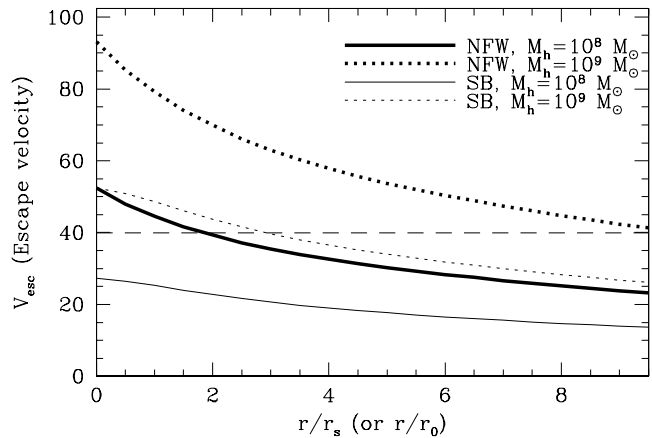,width=8.5cm}
\caption{ 
The radial dependences of the escape velocities ($V_{\rm esc}$)
for four different models:
the SB model with  
$M_{\rm h}=10^8 {\rm M}_{\odot}$ 
and $r_{0}=0.48$ kpc (thick solid), 
the SB one with  
$M_{\rm h}=10^9 {\rm M}_{\odot}$
and $r_{0}=1.28$ kpc (thick dotted), 
the NFW model with
$M_{\rm h}=10^8 {\rm M}_{\odot}$ and $r_{\rm s}=0.16$ kpc (thin solid),
and the NFW one with
$M_{\rm h}=10^9 {\rm M}_{\odot}$ and $r_{\rm s}=0.52$ kpc
(thin dotted).
For comparison, the maximum value of the wind velocity
of AGB stars is shown by a dashed line.
}
\label{Figure. 1} 
\end{figure}

\section{Three advantages of chemical pollution within subhalos}

In the present study, we adopt an  AGB pollution scenario
(e.g., Smith \& Norris 1982)  in
which the second generation of stars originate from AGB ejecta only after
gas ejected from Type II supernovae (SNe II) were removed from
the systems  owing to much more energetic processes of SNe II (with the
wind velocities of an order of $\sim 1000$ km s$^{-1}$).
Progenitor stars of SNe II are considered to have masses 
of $M > 10 {\rm M}_{\odot}$  (e.g., Woosley \& Weaver 1995)
and thus have shorter lifetimes than AGB progenitor stars
with $0.8 < (M/{\rm M}_{\odot}) < 8$
(e.g., van den Hoek \& Groenewegen 1997). Therefore the above assumption
on the  star formation from AGB ejecta only after the completion of SNe II events
is regarded as reasonable. It is expected that the first 
and the second generations
of stars are different with each other 
not in heavier elements (e.g., Fe) but in
CN-processed ones owing to the selective self-enrichment.
Although this scenario has several problems in explaining
{\it quantitatively} the observed abundance pattern (e.g., C and O) in GCs
(e.g., Smith \& Norris 1982),
we adopt this in order to more clearly demonstrate the advantages of
the chemical pollution of GCs within dwarfs embedded in subhalos. 

\subsection{Trapping AGB ejecta}

By investigating whether the terminal (or expansion) velocity ($V_{\rm w}$)
of the stellar wind
of AGB stars can be smaller or larger than the escape velocity
($V_{\rm esc}$) of  a halo (or a GC),
we can discuss whether the ejecta from AGB stars can be trapped
within the GC  for a times scale long enough for star formation
of the second generation of stars (e.g., Smith 1996).
The observed $V_{\rm w}$ ranges roughly from 2 km s$^{-1}$ to 
40 km s$^{-1}$  in Loup et al. (1993) whereas Dupree et al. (1992)
found evidences of AGB winds with $V_{\rm w}=90$ km s$^{-1}$.
In this paper, we discuss whether $V_{\rm esc}$
of a dark matter halo can be smaller
or larger  than  
the possibly maximum value of $V_{\rm w}$ = 40 km s$^{-1}$
(hereafter referred to as $V_{\rm w,max}$).
In order to show the dependences of $V_{\rm esc}$
of low-mass halos with different masses ($M_{\rm h}$) more clearly,
we do not intend to derive $V_{\rm w}$ for  the combined
mass distributions of halos and GCs.

For comparison,  we adopt the two different profiles of dark matter
halos at high redshifts:
(1) The ``SB'' profiles with large cores proposed  by Salucci \& Burkert (2000)
and (2) the ``NFW'' ones   with cuspy cores predicted from the standard cold
dark matter (CDM) cosmogony (Navarro, Frenk, \& White 1996).
The SB profiles consistent with the observed rotation curves of galaxies
are described as: 
\begin{equation}
{\rho}_{\rm sb}(r)=\frac{\rho_{\rm sb,0}}{(r+r_{\rm 0})(r^2+r_{\rm 0})^2},
\end{equation} 
where $\rho_{\rm sb,0}$ and $r_{\rm 0}$ are the central dark matter 
density and the core (scale) radius, respectively.
For the SB profile, the dark matter
core parameters, $\rho_{\rm sb,0}$,  $r_{\rm 0}$,  and $M_{0}$
(where $M_{0}$ is the total dark matter mass within $r_{\rm 0}$)
are not free parameters, and clear correlations are observed between
them (Salucci \& Burkert 2000): 
\begin{equation}
M_{0}=4.3 \times 10^7 {(\frac{r_{\rm 0}}{\rm kpc})}^{7/3} M_{\odot}.
\end{equation} 

 The NFW profile is described as:
\begin{equation}
{\rho}(r)=\frac{\rho_{s}}{(r/r_{\rm s})(1+r/r_{\rm s})^2},
\end{equation}
where $r$,  $\rho_{s}$,  and $r_{\rm s}$ are the distance from the center 
of the cluster, the characteristic  density, and the scale-length of the dark halo, 
respectively.
Fig. 1 shows the radial dependences  of $V_{\rm esc}$ calculated
for the SB model with 
$M_{\rm h}=10^8 {\rm M}_{\odot}$
and $r_{0}=0.48$ kpc, the SB one with 
$M_{\rm h}=10^9 {\rm M}_{\odot}$ 
and $r_{0}=1.28$ kpc, 
the NFW model with 
$M_{\rm h}=10^8 {\rm M}_{\odot}$ and $r_{\rm s}=0.16$ kpc,
and the NFW one with
$M_{\rm h}=10^9 {\rm M}_{\odot}$ and $r_{\rm s}=0.52$ kpc.
These values of the NFW models are derived by using  the lowest
mass model in the NFW with a mass-size scaling of 
$M_{\rm h} \propto  {r_{\rm s}}^2$ reasonable for the CDM halos.
 
As shown in Fig. 1,
$V_{\rm esc}$ 
both for the two NFW models are  significantly larger than $V_{\rm w, max}$
in the nuclear regions with $r/r_{\rm s} < 1$,
which suggests that ejecta of AGB stars can be more effectively 
trapped in nuclear regions of the halos
compared with their outer ones.
The difference between $V_{\rm esc}$ and $V_{\rm w, max}$
at $r/r_{\rm s} < 1$  (or $r/r_{\rm 0} < 1$)
is larger in the more massive model ($M_{\rm h}=10^9 {\rm M}_{\odot}$)
than in the less massive one ($M_{\rm h}=10^8 {\rm M}_{\odot}$) 
both for the NFW and the SB models.
This result implies that ejecta of AGB stars can be more effectively
trapped in more massive halos
and thus that primordial pollution can proceed more efficiently in
more massive halos.
The higher $V_{\rm esc}$ at a given radius for a given $M_{\rm h}$
in the NFW models 
in comparison with the SB models 
suggests that inner density profiles of dark matter halos
can be an important parameter which determines the details of
primordial pollution processes. 

Since the mass densities (${\rho}_{\rm h}$) 
of halos virialized at redshift $z$ are
roughly proportional to $(1+$z$)^3$ (Padmanabhan 1993),
$V_{\rm esc}$ can be larger in halos formed at higher redshifts
for a given halo mass 
(i.e., $V_{\rm esc} \propto {( {M_{\rm h}}^2 {\rho}_{\rm h} )}^{1/6}
\propto {(1+z)}^{1/2}$ for a given $M_{\rm h}$).
This suggests that ejecta of AGB stars can be  more effectively
trapped in halos formed at higher redshifts for a given mass:
Primordial pollution of GCs  can more efficiently proceed in nuclear regions
of halos formed at higher redshifts.

\begin{figure}
\psfig{file=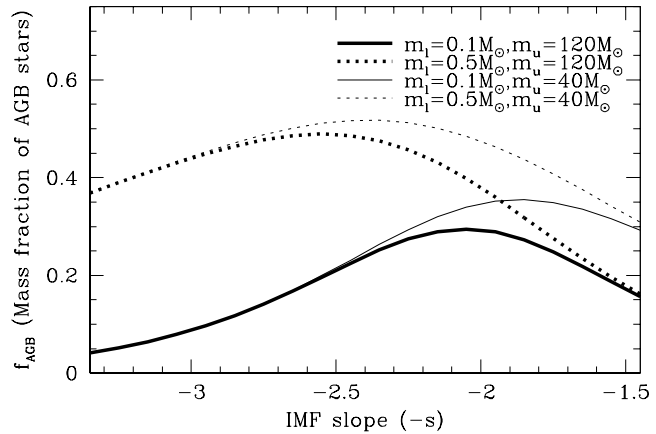,width=8.5cm}
\caption{ 
The dependences of the 
mass fraction of stars with masses ranging from $1 {\rm M}_{\odot}$
to  $8 {\rm M}_{\odot}$ (i.e., progenitors of AGB stars)
on the slopes of IMF (initial mass function) 
with $-3.5 \le -s \le -1.5$ in GCs
for four different models  
with $m_{\rm l}=0.1 {\rm M}_{\odot}$ 
and $m_{\rm u}=120 {\rm M}_{\odot}$ (thick solid), 
with $m_{\rm l}=0.5 {\rm M}_{\odot}$ 
and $m_{\rm u}=120 {\rm M}_{\odot}$ (thick dotted), 
with $m_{\rm l}=0.1 {\rm M}_{\odot}$ 
and $m_{\rm u}=40 {\rm M}_{\odot}$ (thin solid), 
and with $m_{\rm l}=0.5 {\rm M}_{\odot}$ 
and $m_{\rm u}=40 {\rm M}_{\odot}$ (thin dotted). 
If we assume that ejecta of only more massive AGB stars with the masses
of $ 6-7{\rm M}_{\odot}$ can be used for star formation,
$f_{\rm AGB}$ is much smaller than those shown in this figure
(Bekki \& Norris 2005). 
}
\label{Figure. 2} 
\end{figure}

\subsection{Suppression of ram pressure stripping}

Gas within GCs can be stripped if the following inequality
is satisfied:
\begin{equation} 
{\rho}_{\rm ISM} {v_{\rm rel}}^2 > 2\pi G {\Sigma}_{s}
{\Sigma}_{g},
\end{equation} 
where ${\rho}_{\rm ISM}$ is the density of interstellar gas in halos or disks,
$v_{\rm rel}$ is the relative velocity between a GC and the interstellar
gas,  $G$ is the gravitational constant,
${\Sigma}_{s}$ is the gravitational surface mass density,
and ${\Sigma}_{g}$ is the  surface density of intracluster gas
(e.g., ejecta of AGB stars).
By using the formula essentially the same as  the above equation (4),
Frank \& Gisler (1976) showed that ${\rho}_{\rm ISM}$
required for the sweeping of gas 
{\rm within the central regions} of most GCs is $5 \times 10^{-3}$ cm$^{-3}$
for $v_{\rm rel}=200$ km s$^{-1}$,  
${\Sigma}_{s} = 10^4$ ${\rm M}_{\odot}$ pc$^{-2}$,
and a reasonable value of the specific mass loss rate
(used for the estimation of ${\Sigma}_{g}$).
The typical mass and tidal radius ($r_{\rm t}$) of GCs
are $5 \times 10^5 {\rm M}_{\odot}$ and 50 pc, respectively,
(Binney \& Tremaine 1987).
This means that mean ${\Sigma}_{s}$ within $r_{\rm t}$
is by a factor of $\sim 20$
smaller than the above central value and thus  ${\rho}_{\rm ISM}$
required for the sweeping of the outer gas is 
$2.5 \times 10^{-4}$ cm$^{-3}$.

It should be noted here that the derived  values of ${\rho}_{\rm ISM}$
are much smaller than a typical value of the number density of H near
the Sun ($n_{\rm H} \sim 1 {\rm cm}^{-3}$). 
The total mass of dark matter  within  the central 50 pc ($=r_{\rm t}$)  
of the halo in the NFW model with  
$M_{\rm h}=10^9 {\rm M}_{\odot}$ is $2.8 \times 10^6 {\rm M}_{\odot}$,
which means that ${\Sigma}_{s}$ can be significantly  larger
in GCs located in nuclear regions of the halo than in 
isolated GCs. This suggests that ram pressure stripping
is much less effective in sweeping gas of GCs if
GCs are located in nuclear regions of halos.
Although nuclear GCs of halos can keep their gas from been stripped 
by the Galactic halo gas,
the surface gravity from their host halos
appears to be  not strong enough to prevent gas from being swept by
the ISM of the Galactic disk with $n_{\rm H} \sim 1 {\rm cm}^{-3}$.

\subsection{Increase of the total amount of AGB ejecta}

In order to estimate the  mass fraction ($f_{\rm AGB}$)
of AGB progenitor stars with
the masses ranging from $1 {\rm M}_{\odot}$
to  $8 {\rm M}_{\odot}$ in a GC with the total mass of $M_{\rm cl}$,  
we assume an initial mass function
(IMF) in number described
as $\psi (m_{\rm I}) = A{m_{\rm I}}^{-s}$, 
where $m_{\rm I}$ is the initial mass of
each individual star and the slope of $s=2.35$ corresponds to the Salpeter IMF.
The normalization factor $A$ is a function of $M_{\rm cl}$,
$m_{\rm l}$ (lower mass cut-off), and $m_{\rm u}$ (upper one):
\begin{equation}
A=\frac{M_{\rm cl} \times (2-s)}{{m_{\rm u}}^{2-s}-{m_{\rm l}}^{2-s}}.
\end{equation} 
where $m_{\rm l}$ and $m_{\rm u}$ are  regarded as free parameters
in the present study.

Fig. 2 shows the dependences of $f_{\rm AGB}$ on the IMF slope ($s$)
for four different sets of parameters of $m_{\rm l}$ and $m_{\rm u}$.  
$f_{\rm AGB}$ is at most $\sim 0.5$ in these models. The mass fraction
($f_{\rm 2nd}$) of the second  generation of stars 
is described as,
\begin{equation}
f_{\rm 2nd}  \propto {\epsilon}_{\star} \times m_{\rm ej} \times f_{\rm AGB},
\end{equation} 
where ${\epsilon}_{\star}$ is 
the star formation efficiency of star-forming gas
and $m_{\rm ej}$ is the mass ratio  of gas ejected from AGB
stars to initial stellar masses of progenitors of AGB stars.
${\epsilon}_{\star}$ is observed to range from $\sim 0.01$ (Duerr et al.
1982 for the $\lambda$ Sco complex) to $\sim 0.4$ (Wilking \& Lada 1983 for
the $\rho$ Oph cloud).  If we adopt the table value of $m_{\rm ej}$ (=0.46)
for AGB stars with $m_{\rm I}= 1 {\rm M}_{\odot}$
and $ Z=0.001$ in  van den Hoek \& Groenewegen (1997),
we can find that $f_{\rm 2nd}$ is at most 0.12 ($=0.5 \times 0.4 \times 0.46$).

The derived maximum value of $f_{\rm 2nd}=0.12$ is significantly
smaller than 0.5 that is required  for explaining the observed number ratio
of CN-strong stars to CN-weak ones for NGC 6752 in the 
context of the primordial pollution scenario 
(e.g.,  Smith \& Norris 1982).
Although $f_{\rm 2nd}$ required for explaining
abundance inhomogeneity may well be different
between different clusters
(thus smaller $f_{\rm 2nd}$ is still acceptable for some GCs),
the derived small values of $f_{\rm 2nd}$ imply that
if we assume that the  second  generation of stars are formed 
from the first one {\it initially within GCs,}
the primordial pollution scenario can not explain
quantitatively the observed abundance inhomogeneity of GCs.

However, if GCs are located in nuclear regions of dark matter halos
hosting dwarf galaxies,
the ejecta of field stellar populations 
(i.e., major components of the dwarfs) surrounding the nuclear GCs
can be fueled  into the nuclear regions and converted into
new stars there.
If a large amount of gas ejected from the surrounding stellar populations
can be converted into the second generation of stars within GCs,
the mass fraction of the second  generation can be significantly
boosted up: We do not have to assume an extreme set of parameter values 
for IMF and ${\epsilon}_{\star}$ 
in explaining the observed $f_{\rm 2nd}$ 
(See Smith \& Norris 1982 for this problem of unusual IMF required
for the primordial pollution scenario).

\section{Discussions and conclusions}

\subsection{Defunct dwarfs as GC hosts}

A key question in this scenario is whether GC host dwarfs embedded in
dark matter halos can be completely destroyed 
by the strong Galactic tidal field without the central
GCs being destroyed. Recent numerical simulations
have demonstrated that nuclear star clusters can survive from
the strong tidal field of the Galaxy owing to their initial 
compactness whereas the main bodies of (nucleated) dwarfs 
are completely destroyed and dispersed into the Galactic halo
region (e.g., Bekki \& Freeman 2003; Tsuchiya et al. 2003). 
Previous simulations also showed that nucleated dwarfs can be
more efficiently
converted into massive star clusters (i.e., naked nuclei) under
strong tidal field of groups and clusters of galaxies 
for the SB profiles of dark matter halos (Bekki et al. 2003).
These theoretical works therefore strongly suggest that the present
scenario is promising.

Recent abundance studies of stars both for the Galactic halo
and for dwarfs (e.g., dwarf spheroidal, dSph)  in the Local Group have revealed that 
[$\alpha$/Fe] ratios of most stars in the dwarfs are 
generally lower than similar metallicity Galactic halo stars 
(e.g., Venn et al. 2004).
The higher [$\alpha$/Fe] ratios  of the Galactic halo can be due
to {\it very early merging of low-mass dwarf galaxies}, which
were destroyed to form the Galactic halo without efficient 
chemical enrichment of stars
(Venn et al. 2004). 
We propose  that these dwarfs destroyed in the very early history
of the Galaxy
(i.e., defunct dwarfs) are host galaxies of the Galactic halo
GCs with abundance inhomogeneity.
This proposal is consistent with the observed higher [$\alpha$/Fe]
ratios in the Galactic GCs (e.g., Freeman 1993).

\subsection{Diversity in primordial pollution processes}

Smith (1987) divided GCs into three classes according to
(1) whether star-to-star abundance inhomogeneity can be
seen for Fe-peak element and (2) whether it can be seen
in CN abundance (i.e., whether GCs show bimodal CN distributions).
It was found in Smith (1987) that (1) only $\omega$ Cen and M22 show 
inhomogeneity in Fe-peak elements,
(2) GCs with no bimodal CN distributions have lower metallicities
of $-2.2 \le {\rm [Fe/H]} \le -1.6$ (e.g., M92 and M15),
and (3) GCs with bimodal CN distributions have
higher metallicities of ${\rm [Fe/H]} > -1.6$
(e.g., NGC 6752 and 47 Tuc).
However it remains unclear why  $\omega$ Cen 
shows such
inhomogeneity in Fe-peak element.

The present study can provide the 
following  answer for the above question,
by assuming that the observed inhomogeneity is solely due to
the primordial pollution processes of GCs.
$\omega$ Cen was formed in the nuclear region of a massive subhalo 
($\sim 5 \times 10^8 {\rm M}_{\odot}$) that
was virialized at very high redshift ($z \sim 15$)
and was more massive than any other subhalos hosting the Galactic GCs
at that redshift.
This massive halo 
with a higher mass density 
(due to earlier virialization) could not be soon destroyed by the proto-Galactic
tidal field because of its stronger self-gravity,
and consequently star formation could continue for a longer time scale
in its nuclear region.
As a result of this, not only AGB ejecta but also some fraction
of metal-enriched
gas from Type II and Type Ia 
supernovae  could  be finally recycled and converted
into new generations of stars within the $\omega$ Cen's host halo.
The initially large apocenter distance of its orbit,
for which a longer time scale of dynamical friction is 
required for the $\omega$ Cen's host to reach the Galactic central
region,  can be also
responsible for the  prolonged star formation activity
(Bekki \& Freeman 2003).

Less massive halos 
containing  GCs (e.g., NGC 6752) other than $\omega$ Cen
had shallower gravitational potential and
were more easily destroyed by the tidal field of the proto-Galaxy
in their earlier histories.
Only ejecta from more massive AGB stars with smaller wind
velocities (an order of $\sim 10$ km s$^{-1}$)
therefore could be converted into new stars 
to become the second generation of stars in GCs  without
efficient chemical enrichment 
(i.e.,  without significant metallicity spread) due to SNe II with very large
wind velocities (an order of $\sim 1000$ km s$^{-1}$).
Less massive halos with smaller pericenter and apocenter distances of their orbits 
would not have experienced efficient chemical pollution of GCs 
owing to more rapid destruction of the halos.
Thus $\omega$ Cen  can 
show abundance inhomogeneity both in heavier elements (e.g., Fe) and in
light ones (e.g., C, N, and O) whereas other GCs 
can show  abundance inhomogeneity only in light ones.

\subsection{External pollution scenario}

We have shown that the AGB ejecta can be retained 
in the central regions of their host dwarfs and consequently
converted into stars that finally become the second generation of
stars in GCs. 
We also have suggested  that AGB ejecta of field  stellar populations
surrounding the nuclear GCs can be converted into
stars to become the second generation of stars in GCs. 
Accordingly, it would be reasonable to say that ``external pollution''
(or modified version of the original primordial pollution scenario) 
rather than ``self-pollution'' is a more reasonable jargon 
that denotes the chemical pollution processes of GCs in 
the central regions of dwarfs.
This external pollution scenario may well  explain
not only the observed helium overabundance of $\omega$ Cen
(Bedin et al. 2004)
but also the large fraction of CN-strong stars in GCs
with abundance inhomogeneity (Bekki \& Norris 2005). 

One of potential  problems of this external scenario 
is that the physical mechanism for the selective pollution
by AGB stars (not by SNe II) is not clearly understood.
In order to discuss this point, we plan to investigate
gas dynamics within the central $0.1-100$pc of dwarfs at
very high redshifts by using chemodynamical simulations
combined with the latest results of AGB yields derived by
Campbell et al. (2005). It is also our future study
to investigate whether this scenario  can explain 
the observed O-Na and Mg-Al anticorrelations of GCs.
Smith \& Norris (1982) suggested that if both the first and the second
generations of stars are formed from ejecta of AGB stars with different
mases,  the observed CN-bimodality
in GCs can be explained. This modified primordial (and external)
pollution scenario will be addressed by our future chemodynamical simulations
in a more quantitatively.

Theoretical studies based on  cosmological
simulations have just started extensive investigation
on structural and kinematical properties of GC systems
of galaxies in  a $\Lambda$CDM Universe (e.g., Bekki 2005; Yahagi \& Bekki 2005).
These Mpc-scale simulations, combined with sub-pc scale ones on
star formation within GCs, will provide more robust predictions
on spatial distributions and kinematics of the Galactic GCs with
different past histories of primordial pollution processes
and thus be compared with corresponding
observations (e.g., Carretta 2005).

\section*{Acknowledgments}
We are  grateful to the anonymous referee for valuable comments,
which contribute to improve the present paper.
K.B. acknowledges the financial support of the Australian Research 
Council throughout the course of this work.

\end{document}